\documentclass{kapproc} 

\usepackage{t1enc}

\usepackage{procps} 

\usepackage[dvips]{graphicx}

\upperandlowercase

\kluwerbib 

\begin{document}

\articletitle{Dust attenuation and star formation in the nearby universe: the
ultraviolet and far-infrared points of view}


\author{V\'eronique Buat}
\affil{Laboratoire d'Astrophysique de Marseille, France}
\email{veronique.buat@oamp.fr}

\author{Jorge Iglesias-P\'aramo}
\affil{Laboratoire d'Astrophysique de Marseille, France}

\email{jorge.iglesias@oamp.fr}

\begin{abstract}
We make use of the on-going All Imaging Survey of the UV GALEX satellite
cross-correlated with the IRAS all sky survey to build samples of galaxies
truly selected in far-infrared or in ultraviolet. We discuss the amount of dust 
 attenuation and the star formation rates for these galaxies and
compare the properties of the galaxies selected in FIR or in UV. 
\end{abstract}


\section*{Introduction}
The measure of the star formation rate (SFR) in galaxies is based on the
analysis of the emission from young stars which escapes the galaxies or is
absorbed and  re-emitted by the gas or the dust. In this paper, we
will focus on UV and IR emissions. These emissions are closely related by the
energetic budget at work in star forming galaxies: the UV light which does not
escape the galaxy   is absorbed by the interstellar dust and re-emitted in the
far-IR. Therefore both emissions originate from the same stellar populations
and their comparison is a powerful tracer of the dust attenuation (Buat \& Xu
\cite{buatxu},Gordon et al. \cite{gordonetal}). They are also
closely related to the recent star formation rate over similar timescales
(Buat \& Xu \cite{buatxu},Kennicutt \cite{kennicutt}). In this paper we will
combine the new GALEX data together with the existing far-IR data from IRAS. 

\section{The data}
We have worked on 600 deg$^2$ observed by GALEX in NUV( $\lambda = 231$ nm
and FUV ($\lambda =  153 $ nm)  to build two samples of
galaxies. The first one, called the {\sl NUV selected sample} includes all the
galaxies brighter than m(NUV) = 16 mag (AB scale), among the 88 selected
galaxies (excluding ellipticals and active galaxies) only 3 are not detected
by IRAS at 60 $\mu m$. The second sample, called the {\sl FIR selected sample}
is based on the IRAS PSCz (Saunders et al. \cite{saundersetal}): 118 galaxies
from this catalog lie within our GALEX fields, only 1 is not detected in NUV. 

\section{Dust attenuation in the nearby universe}

\paragraph{Mean values of the dust attenuation}
For both samples we measure the dust attenuation using the F(dust)/F(NUV)
ratio. This ratio is a quantitative measure of the dust attenuation at UV
wavelengths   (e.g. Buat \& Xu \cite{buatxu}, Gordon et al.
\cite{gordonetal}). The formulae used here
are  adapted to the  GALEX 
bands (Buat et al. 2004, ApJL, in press). The total (8-1000 $\mu m$) dust
emission is calculated from the fluxes at 60 and 100 $\mu m$ using the Dale et
al. (\cite{daleetal}) recipe. \\
A moderate attenuation is found in the NUV selected sample with
$0.8^{-0.3}_{+0.4}$  mag in NUV and  and $1.1^{-0.4}_{+0.5}$ mag in FUV. As
expected the dust attenuation is higher in the FIR selected sample with
$2.1^{-0.8}_{+1.2}$ mag in NUV and $2.9^{-1.2}_{+1.2}$ mag in FUV.\\
A comparison of the FIR and UV luminosity densities from IRAS (Saunders et al.
\cite{saundersetal2}) and GALEX (Thilker et al. 2004, ApJL, in press) leads to
a mean dust attenuation at z$\sim$0 of 1.1 mag in NUV and 1.6 mag in FUV: the
nearby universe is not very obscured.

 \begin{figure}
\includegraphics[angle=-90]{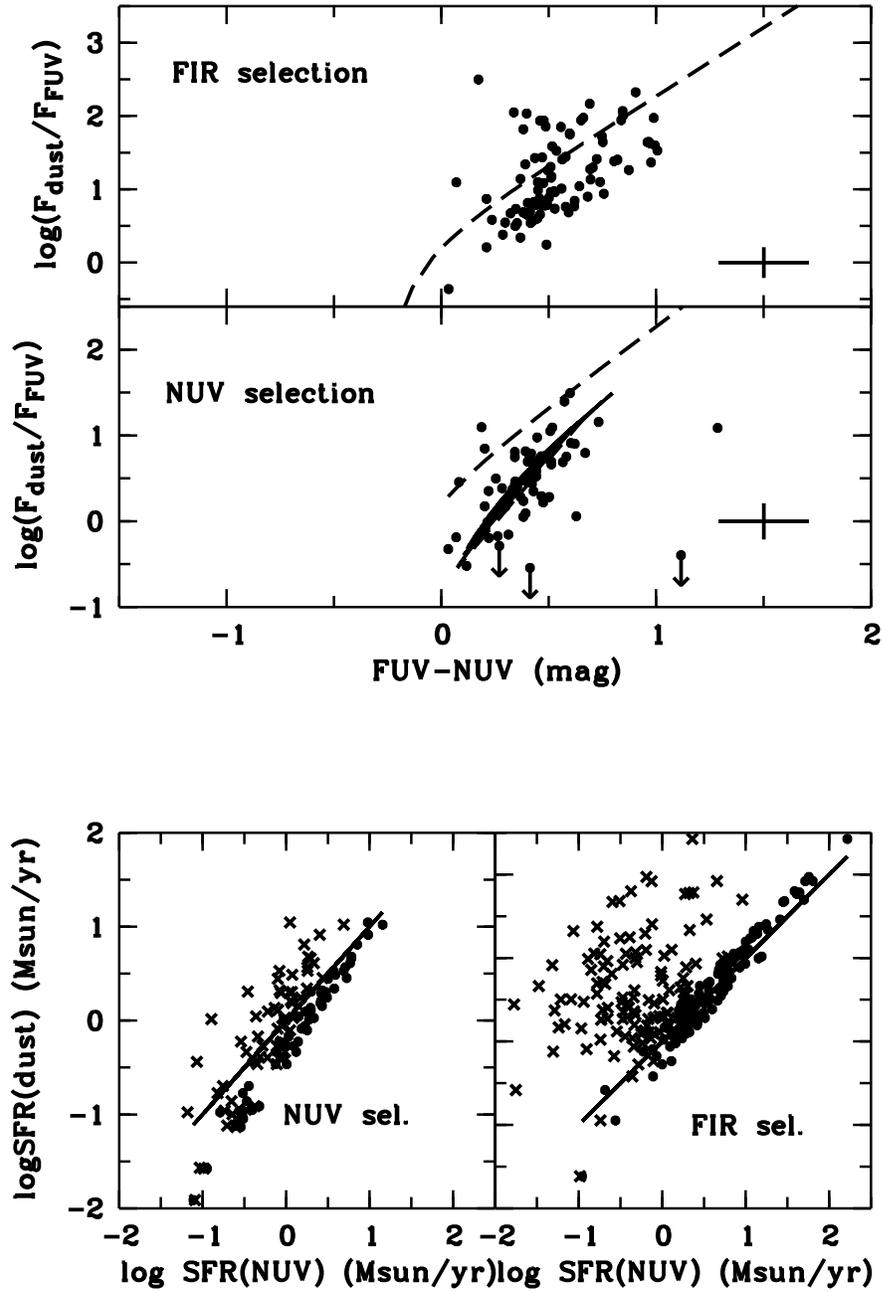}
\caption{{\bf top}: log($F(dust)/F(FUV)$) against the FUV-NUV color for NUV
and FIR
selected samples. The dashed line is the mean relation expected for starburst
galaxies, the  solid line is the locus of Kong et al. models
with b=0.25 (see text).
{\bf bottom}: SFR  deduced from the dust luminosity versus the SFR deduced from
the NUV luminosity directly observed (crosses) and corrected for dust
attenuation (points) for the NUV selected sample (left panel) and the FIR
selected one (right panel). The  lines correspond to equal quantities on
both axes}. 
\end{figure}
\paragraph{The "IRX-$\beta$" relation}
From their study of starburst galaxies Meurer and coll. (e.g. Meurer et al.
\cite{meureretal}) have found a strong correlation between the slope $\beta$
of the UV continuum (fitted as a power law $F(\lambda)\propto \lambda^{\beta}$
between 1200 and 2500 \AA) and F(dust)/F(FUV). They concluded that
$\beta$ is a reliable tracer of the dust attenuation in galaxies. However
further works showed that this law is not universal and that ultra-luminous
infrared galaxies (Goldader et al. \cite{goldaderetal}) and normal spiral
galaxies (Bell \cite{bell}) do not follow the relation found for starbursts.\\
The FUV-NUV color from the GALEX observation is directly linked to $\beta$
(Kong et al. \cite{kongetal}). We have plotted this color versus
F(dust)/F(FUV) for our two samples of galaxies in Fig. 1 and compared to the
predictions for starburst galaxies.
Obviously $\beta$ is not a reliable tracer of the dust attenuation in our NUV
and FIR selected samples. Moreover, different behaviors are found within both
samples: whereas the FIR selected galaxies spread over a large area of the
diagram most of the NUV selected galaxies lie below the starburst relation.
This may be due to age effects in the stellar populations: assuming  a b
parameter 
(defined as the ratio of the  recent SFR divided by the past   averaged  one)
equal to 0.25
leads to a reasonable fit using the models of Kong et al. \cite{kongetal}. 
\section{The measure of the star formation rate}
UV and total dust  emissions can be calibrated in a recent star formation rate
assuming a recent (over $\sim 10^8$ years) star formation history and an
initial mass
function (Buat \& Xu \cite{buatxu}, Kennicutt \cite{kennicutt}). When using
the dust luminosity one must add an additional hypothesis about the absorption 
of the stellar light by the dust. The classical hypothesis (also made in this
work) is that all the stellar light from the young stars is absorbed by the
dust (Kennicutt \cite{kennicutt}). In this paper we also assume a constant SFR
over $10^8$ years, a Salpeter IMF. Using Starburst99 synthesis models we
obtain:\\
$log (L_{NUV}) (L\odot) = 9.73+log(SFR) (M\odot yr^{-1})$ and\\
$log (L_{dust}) (L\odot) = 10.168+log(SFR) (M\odot yr^{-1})$.\\
In Fig 1 are plotted the SFR estimated from the NUV luminosity against the SFR
from the dust luminosity. In both samples, the {\it observed} NUV luminosity
strongly under-estimates the SFR the effect being worse for the FIR selected
sample. When the NUV selected sample is corrected for dust attenuation the
agreement is very good (as expected) between both estimates of the SFR in the
FIR selected sample since we are dominated in each case by the dust
emission. Conversely in the NUV selected sample the SFR estimated from the
dust luminosity alone is found to underestimate the SFR as compared to the NUV
corrected luminosity. The discrepancy increases towards the low SFRs to reach
a factor 3 for SFR of $\sim$ 0.3 M$\odot~ yr^{-1}$. Therefore, using the dust
emission alone to measure the total SFR in all galaxies can be misleading, the
best way being to combine UV and IR emissions to estimate reliable SFRs.


\begin{chapthebibliography}{1}
\bibitem[2002]{bell}
Bell, E. 2002, {\it ApJ} 577, 150
\bibitem[1996]{buatxu}
Buat, V., Xu, C. 1996, {\it A \& A}, 306, 61
\bibitem[2001]{daleetal}
Dale, D., Helou, G., Contursi, A., Silbermann, N., Kolhatkar, S. 2001, {\it
ApJ}, 549, 215
\bibitem[2002]{goldaderetal} 
Goldader,J., D., Meurer, G., Heckman,
T., M., Seibert, M., Sanders, D. B., Calzetti, D., Steidel, C., C. 2002, {\it
ApJ} 
568, 651
\bibitem[2000]{gordonetal}
Gordon, K, Clayton, G., Witt, A., Misselt, K 2000, {\it ApJ} 533, 236
\bibitem[1998]{kennicutt}
Kennicutt, R. 1998, {\it ARAA} 36, 189
\bibitem[2004]{kongetal}
Kong, X., Charlot, S., Brinchmann, J., Fall, M. 2004,  {\it MNRAS} 349. 769
\bibitem[1999]{meureretal} 
Meurer, G.~R., Heckman, T.~M., \& Calzetti, D. 1999, {\it ApJ}, 521, 64 
\bibitem[2000]{saundersetal}
Saunders, W. et al. 2000, {\it MNRAS} 317, 55
\bibitem[1990]{saundersetal2}
Saunders, W., Rowan-Robinson, M., Lawrence, A. et al. 1990, {\it MNRAS} 242,
318
\end{chapthebibliography}
\end{document}